\documentstyle[aps,preprint,pra,graphicx,url]{revtex}
\begin{document}
\newcommand{\etal}{{\em et al.}\/}
\newcommand{\IP}{inner polarization}
\newcommand{\IPF}{\IP\ function}
\newcommand{\IPFs}{\IP\ functions}
\newcommand{\auth}[2]{#1 #2, }
\newcommand{\oneauth}[2]{#1 #2,}
\newcommand{\twoauth}[4]{#1 #2 and #3 #4,}
\newcommand{\andauth}[2]{and #1 #2, }
\newcommand{\book}[4]{{\it #1} (#2, #3, #4)}
\newcommand{\jcite}[4]{#1 #2(#4)#3}
\newcommand{\et}{ and }
\newcommand{\erratum}[3]{\jcite{erratum}{#1}{#2}{#3}}
\newcommand{\inpress}[1]{{\it #1}}
\newcommand{\inbook}[5]{In {\it #1}; #2; #3: #4, #5}
\newcommand{\JCP}[3]{\jcite{J. Chem. Phys.}{#1}{#2}{#3}}
\newcommand{\jms}[3]{\jcite{J. Mol. Spectrosc.}{#1}{#2}{#3}}
\newcommand{\jmsp}[3]{\jcite{J. Mol. Spectrosc.}{#1}{#2}{#3}}
\newcommand{\jmstr}[3]{\jcite{J. Mol. Struct.}{#1}{#2}{#3}}
\newcommand{\cpl}[3]{\jcite{Chem. Phys. Lett.}{#1}{#2}{#3}}
\newcommand{\cp}[3]{\jcite{Chem. Phys.}{#1}{#2}{#3}}
\newcommand{\pr}[3]{\jcite{Phys. Rev.}{#1}{#2}{#3}}
\newcommand{\jpc}[3]{\jcite{J. Phys. Chem.}{#1}{#2}{#3}}
\newcommand{\jpcA}[3]{\jcite{J. Phys. Chem. A}{#1}{#2}{#3}}
\newcommand{\jpca}[3]{\jcite{J. Phys. Chem. A}{#1}{#2}{#3}}
\newcommand{\jpcB}[3]{\jcite{J. Phys. Chem. B}{#1}{#2}{#3}}
\newcommand{\PRA}[3]{\jcite{Phys. Rev. A}{#1}{#2}{#3}}
\newcommand{\PRB}[3]{\jcite{Phys. Rev. B}{#1}{#2}{#3}}
\newcommand{\jcc}[3]{\jcite{J. Comput. Chem.}{#1}{#2}{#3}}
\newcommand{\molphys}[3]{\jcite{Mol. Phys.}{#1}{#2}{#3}}
\newcommand{\mph}[3]{\jcite{Mol. Phys.}{#1}{#2}{#3}}
\newcommand{\APJ}[3]{\jcite{Astrophys. J.}{#1}{#2}{#3}}
\newcommand{\cpc}[3]{\jcite{Comput. Phys. Commun.}{#1}{#2}{#3}}
\newcommand{\jcsfii}[3]{\jcite{J. Chem. Soc. Faraday Trans. II}{#1}{#2}{#3}}
\newcommand{\prsa}[3]{\jcite{Proc. Royal Soc. A}{#1}{#2}{#3}}
\newcommand{\jacs}[3]{\jcite{J. Am. Chem. Soc.}{#1}{#2}{#3}}
\newcommand{\ijqcs}[3]{\jcite{Int. J. Quantum Chem. Symp.}{#1}{#2}{#3}}
\newcommand{\ijqc}[3]{\jcite{Int. J. Quantum Chem.}{#1}{#2}{#3}}
\newcommand{\spa}[3]{\jcite{Spectrochim. Acta A}{#1}{#2}{#3}}
\newcommand{\tca}[3]{\jcite{Theor. Chem. Acc.}{#1}{#2}{#3}}
\newcommand{\tcaold}[3]{\jcite{Theor. Chim. Acta}{#1}{#2}{#3}}
\newcommand{\jpcrd}[3]{\jcite{J. Phys. Chem. Ref. Data}{#1}{#2}{#3}}
\newcommand{\science}[3]{\jcite{Science}{#1}{#2}{#3}}
\newcommand{\CR}[3]{\jcite{Chem. Rev.}{#1}{#2}{#3}}

\draft
\title{A fully {\it ab initio} potential curve of near-spectroscopic
quality for OH$^-$ ion: importance
of connected quadruple excitations and scalar relativistic effects}
\author{Jan M.L. Martin*}
\address{Department of Organic Chemistry,
Kimmelman Building, Room 262,
Weizmann Institute of Science,
IL-76100 Re\d{h}ovot, Israel. {\rm E-mail:} {\tt comartin@wicc.weizmann.ac.il}
}
\date{Special issue of {\it Spectrochimica Acta A}: Received March 6, 2000; In final form March 16, 2000}
\maketitle
\begin{abstract}
A benchmark study has been carried out on the ground-state potential 
curve of the hydroxyl anion, OH$^{-}$, including detailed calibration
of both the 1-particle and n-particle basis sets. The CCSD(T) basis 
set limit overestimates $\omega_e$ by about 10 cm$^{-1}$, which is only 
remedied by inclusion of 
connected quadruple excitations in the coupled cluster expansion
--- or, equivalently, the inclusion of the $2\pi$ orbitals in the
active space of a multireference calculation. Upon inclusion of 
scalar relativistic effects (-3 cm$^{-1}$ on $\omega_e$), a potential
curve of spectroscopic quality (sub-cm$^{-1}$ accuracy) is obtained.
Our best computed EA(OH), 1.828 eV, agrees to three decimal
places with the best available experimental value. Our best computed
dissociation energies, $D_0$(OH$^-$)=4.7796 eV and $D_0$(OH)=4.4124 eV,
suggest that the experimental $D_0$(OH)=4.392 eV may possibly be about 
0.02 eV too low.
\end{abstract}

\section{Introduction}

Molecular anions play an important role in the chemistry of the interstellar
medium\cite{r1}, of carbon stars\cite{r2}, and the Earth's ionosphere\cite{r3}.
As pointed out in Ref.\cite{Lee99}, the presence of anions in the
interstellar medium may have profound consequences for our understanding
of the interstellar processing of the biogenic elements (see e.g. Ref.\cite{All97}
and references therein).

Yet as judged from the number of entries in the compilations of Huber and
Herzberg\cite{Hub79} (for diatomics) and of Jacox\cite{Jacox} (for 
polyatomics), 
high- or even medium-resolution spectroscopic data for anions are relatively scarce compared
to the amount of data available for neutral or even cationic species: 
in the 1992 review of Hirota\cite{Hirota} on spectroscopy of ions,
only 13 molecular anions were listed in Table VII, compared to 4 1/2 pages
worth of entries for cations. (Early reviews of anion spectroscopy
are found in Refs.\cite{Amano,Saykally}, while ab initio studies
of structure and spectroscopy of anions were reviewed fairly recently
by Botschwina and coworkers\cite{Bot95}.)
Some of the reasons
for this paucity are discussed in the introductions to Refs.\cite{Lee97,Lee99}.

One such species is the hydroxyl anion, OH$^-$. By means of velocity
modulation spectroscopy\cite{velmod}, high-resolution fundamentals
were obtained\cite{Ros86,Reh86}
 for three isotopomers, namely $^{16}$OH$^-$, $^{16}$OD$^-$,
and $^{18}$OH$^-$; in addition, some pure rotational transitions have
been observed\cite{Oka86}.
Lineberger and coworkers\cite{Sch82} earlier obtained some rotational
data in the course of an electron photodetachment study, and obtained
precise electron affinities (EAs)  of 14741.03(17) and 14723.92(30) cm$^{-1}$, 
respectively, for OH and OD. Very recently, the same group re-measured\cite{Smi97}
EA(OH) and obtained essentially the same value but with a higher precision,
14741.02(3) cm$^{-1}$.

The spectroscopic constants of OH$^-$ were previously the subject of 
ab initio studies, notably by Werner et al.\cite{Wer83} using multireference
configuration interaction (MRCI) methods, and recently by Lee and
Dateo (LD)\cite{Lee97} using coupled cluster theory with basis sets as large as
$[7s6p5d4f3g2h/6s5p4d3f2g]$.

The LD paper is particularly relevant here. The CCSD(T) (coupled cluster
with all single and double substitutions\cite{Pur82} and a quasiperturbative
treatment for triple excitations\cite{Rag89}) method, in combination with
basis sets of at least $spdfg$ quality and including an account for 
inner-shell correlation, can routinely predict vibrational band origins
of small polyatomic molecules with a mean absolute error on the order of
a few cm$^{-1}$ (e.g. for C$_2$H$_2$\cite{c2h2}, SO$_2$\cite{so2}). Yet
while LD found very good agreement between their computed 
CCSD(T)/[6s5p4d3f2g/5s4p3d2f] spectroscopic constants and available
experimental data, consideration of further basis set expansion and of
inner-shell correlation effects leads to a predicted fundamental $\nu$
at the CCSD(T) basis set limit of 3566.2$\pm$1 cm$^{-1}$, about 11 cm$^{-1}$
higher than the experimental results\cite{Ros86} of 3555.6057(22) cm$^{-1}$,
where the uncertainty in parentheses represents two standard deviations.

In a recent benchmark study\cite{ch} on the ground-state potential
curves of the first-row diatomic hydrides using both CCSD(T) and FCI
(full configuration interaction) methods, the author found that CCSD(T)
has a systematic tendency to overestimate harmonic frequencies of A--H
stretching frequencies by on the order of 6 cm$^{-1}$. Even so, the
discrepancy seen by LD is a bit out of the ordinary, and the question
arises as to what level of theory is required to obtain `the right result for
the right reason' in this case. 

In the present work, we shall show that the discrepancy between the CCSD(T)
basis set limit and Nature is mostly due to two factors: (a) neglect of the
effect of connected quadruple excitations, and (b) neglect of scalar 
relativistic effects. When these are properly accounted for, the available
vibrational transitions can be reproduced to within a fraction of a cm$^{-1}$
from the computed potential curve. In the context of the present Special Issue,
this will also serve as an illustrative example of the type of accuracy that
can be achieved for small systems with the present state of the art. Predicted
band origins for higher vibrational levels (and `hot bands') may assist
future experimental work on this system. Finally,
as by-products of our analysis, we will show that the electron affinity of
OH can be reproduced to very high accuracy, and tentatively propose a slight upward
revision of the dissociation energy of neutral hydroxyl radical, OH.

\section{Computational methods}

The coupled cluster, multireference averaged coupled pair functional (ACPF)\cite{Gda88}, and full CI calculations were carried out using MOLPRO 98.1\cite{molpro}
running on DEC/Compaq Alpha workstations in our laboratory, and on the
SGI Origin 2000 of the Faculty of Chemistry. Full CCSDT (coupled cluster theory
with all connected single, double and triple excitations\cite{ccsdt}) and 
CCSD(TQ) (CCSD with quasiperturbative corrections for triple and 
quadruple excitations\cite{ccsdparTQ}) 
calculations were carried out using ACES II\cite{aces} on a DEC Alpha workstation.

Correlation consistent basis sets due to Dunning and coworkers\cite{Dun89,ccecc}
were used throughout. Since the system under consideration is anionic, the
regular cc-pV$n$Z (correlation consistent polarized valence $n$-tuple zeta,
or V$n$Z for short)
basis sets will be inadequate. We have considered both 
the aug-cc-pV$n$Z (augmented
correlation consistent, or AV$n$Z for short) basis sets\cite{Ken92} in which one low-exponent function
of each angular momentum is added to both the oxygen and hydrogen
basis sets, as well as the aug$'$-cc-pV$n$Z basis sets\cite{Del93} 
in which the addition
is not made to the hydrogen basis set. In addition we consider both uncontracted
versions of the same basis sets (denoted by the suffix "uc") and 
the aug-cc-pCV$n$Z basis sets\cite{cvqz} (ACV$n$Z) 
which include added core-valence
correlation functions.  The largest basis sets considered in this work,
aug-cc-pV6Z and aug-cc-pCV5Z, are of [8s7p6d5f4g3h2i/7s6p5d4f3g2h] and
[11s10p8d6f4g2h/6s5p4d3f2g] quality, respectively.

The multireference ACPF calculations were carried out from a CASSCF
(complete active space SCF) reference wave function with an active 
space consisting of the valence $(2\sigma)(3\sigma)(1\pi)(4\sigma)$
orbitals as well as the $(2\pi)$ Rydberg orbitals: this is denoted
CAS(8/7)-ACPF (i,e, 8 electrons in 7 orbitals). While the inclusion
of the $(2\pi)$ orbitals is essential (see below), the inclusion of 
the
$(5\sigma)$ Rydberg orbital (i.e., CAS(8/8)-ACPF) was considered and
found to affect computed properties negligibly. In addition, some
exploratory CAS-AQCC (averaged
quadratic coupled cluster\cite{aqcc}) calculations were also 
carried out.

Scalar relativistic effects were computed as expectation values of the
one-electron Darwin and mass-velocity operators\cite{Cow76,Mar83} for
the ACPF wave functions.

The energy was evaluated at 21 points around $r_e$, with a spacing
of 0.01 \AA. (All energies were converged to 10$^{-12}$ hartree, or
wherever possible to 10$^{-13}$ hartree.) A polynomial in $(r-r_e)/r_e$
of degree
8 or 9 (the latter if an F-test revealed an acceptable statistical
significance for the nonic term) was fitted to the energies.
Using the procedure detailed in Ref.\cite{ch}, the Dunham series\cite{Dun32} thus
obtained was transformed by derivative matching into a variable-beta Morse
(VBM) potential\cite{Cox92} 
\begin{equation}
V_c = D_e \left(1-\exp[-z (1 + b_1 z + b_2 z^2 + \ldots+b_6 z^6)]\right)^2
\end{equation}
in which $z\equiv \beta (r-r_e)/r_e$, $D_e$ is the (computed or observed)
dissociation energy, and $\beta$ is an adjustable parameter
related to that in the Morse function. Analysis of this function was
then carried out in two different manners: (a) analytic differentiation
with respect to $(r-r_e)/r_e$
up to the 12th derivative followed by a 12th-order Dunham
analysis using an adaptation of the ACET program of Ogilvie\cite{acet}; and 
(b) numerical integration of the one-dimensional Schr\"odinger
equation using the algorithm of Balint-Kurti et al.\cite{balint}, on a grid of
512 points over the interval 0.5$a_0$---5$a_0$. As expected, differences between
vibrational energies obtained using both methods are negligible up to the
seventh vibrational quantum, and still no larger than 0.4 cm$^{-1}$ for the
tenth vibrational quantum.

\section{Results and discussion}

\subsection{$n$-particle calibration}

The largest basis set in which we were able to obtain a full CI 
potential curve was cc-pVDZ+sp(O), which means the standard cc-pVDZ
basis set with the diffuse $s$ and $p$ function from aug-cc-pVDZ added
to oxygen. A comparison of computed properties for OH$^-$ with different electron
correlation methods is given in Table \ref{npart}, while their errors in the total
energy relative to full CI are plotted in Figure 1.

It is immediately seen that CCSD(T) exaggerates the curvature of the potential
surface, overestimating $\omega_e$ by 10 cm$^{-1}$. In addition, it 
underestimates the bond length by about 0.0006 \AA. These are slightly more 
pronounced variations on trends previously seen\cite{ch} for the OH radical.

The problem does not reside in CCSD(T)'s quasiperturbative treatment of
triple excitations: performing a full CCSDT calculation instead lowers
$\omega_e$ by only 1.7 cm$^{-1}$ and lengthens the bond by less than 
0.0001 \AA. Quasiperturbative inclusion of connected quadruple excitations,
however, using the CCSD(TQ) method, lowers $\omega_e$ by 8.5 cm$^{-1}$
relative to CCSD(T), and slightly lengthens the bond, by 0.00025 \AA.
(Essentially the same result was obtained by means of the CCSD+TQ*
method\cite{ccsd+tq*}, which differs from CCSD(TQ) in a small sixth-order
term $E_{6TT}$.)
No CCSDT(Q) code was available to the author: approximating the CCSDT(Q)
energy by the expression $E[CCSDT(Q)]\approx E[CCSDT]+ E[CCSD(TQ)] - E[CC5SD(T)] = E[CCSDT] + E_{5QQ} + E_{5QT}$, we obtain a potential curve in fairly good agreement with full CI.

What is the source of the importance of connected quadruple excitations
in this case? Analysis of the FCI wave function reveals prominent contributions
to the wave function from $(1\pi)^4(2\pi)^0\rightarrow(1\pi)^2(2\pi)^2$ double excitations;
while the $(2\pi)$ orbitals are LUMO+2 and LUMO+3 rather than LUMO, a large
portion of them sits in the same spatial region as 
the occupied $(1\pi)$ orbitals. In any proper multireference
treatment, the aforementioned  excitations would be in the zero-order wave
function: obviously, the space of all double excitations therefrom would 
also entail quadruple excitations with respect to the Hartree-Fock 
reference, including a connected component.

Since the basis set sizes for which we can hope to perform CCSDT(Q) or similar
calculations on this system are quite limited, we considered multireference
methods, specifically ACPF from a $[(2\sigma)(3\sigma)(4\sigma)(1\pi)(2\pi)]^8$
reference space (denoted ACPF(8/7) further on). As might be expected, the
computed properties are in very close agreement with FCI, except for $\omega_e$
being 1.5 cm$^{-1}$ too high. AQCC(8/7) does not appear to represent a further
improvement, and adding the $(5\sigma)$ orbital to the ACPF reference space
(i.e. ACPF(8/8)) affects properties only marginally.

\subsection{1-particle basis set calibration}

All relevant results are collected in Table \ref{1part}.
Basis set convergence in this system was previously studied in some detail
by LD at the CCSD(T) level. Among other things, they noted
that $\omega_e$ still changes by 4 cm$^{-1}$ upon expanding the basis
set from aug-cc-pVQZ to aug-cc-pV5Z. They suggested that $\omega_e$
then should be converged to about 1 cm$^{-1}$; this statement is 
corroborated by the CCSD(T)/aug-cc-pV6Z results. 

Since the negative charge resides almost exclusively on the oxygen,
the temptation exists to use aug$'$-cc-pV$n$Z basis sets, i.e. to
apply aug-cc-pV$n$Z only to the oxygen atom but use a regular cc-pV$n$Z
basis set on hydrogen. For $n$=T, this results in fact in a difference
of 10 cm$^{-1}$ on $\omega_e$, but the gap narrows as $n$ increases.
Yet extrapolation suggests convergence of the computed fundamental
to a value about 1 cm$^{-1}$ higher than the aug-cc-pV$n$Z curve. 

For the AV$n$Z and A'V$n$Z basis sets ($n$=T,Q), 
the CAS(8/7)-ACPF approach systematically lowers harmonic frequencies by
about 8 cm$^{-1}$ compared to CCSD(T); for the fundamental the difference
is even slightly larger (9.5 cm$^{-1}$).  Interestingly, this difference
decreases for $n$=5. 

It was noted previously\cite{ch} that the higher anharmonicity constants
exhibit rather greater basis set dependence than one might reasonably
have expected, and that this sensitivity is greatly reduced if uncontracted
basis sets are employed (which have greater radial flexibility). The same
phenomenon is seen here. 

In agreement with previous observations by LD,
inner-shell correlation reduces the bond
lengthen slightly, and increases $\omega_e$ by 5--6 cm$^{-1}$. This occurs
both at the CCSD(T) and the CAS(8/7)-ACPF levels. 

\subsection{Additional corrections and best estimate}

At our highest level of theory so far, 
namely CAS(8/7)-ACPF(all)/ACV5Z, $\nu$ is
predicted to be 3559.3 cm$^{-1}$, still several cm$^{-1}$ higher than 
experiment. 
The effects of further basis set improvement can be gauged from the
difference between CCSD(T)/AV6Z and CCSD(T)/AV5Z results: one notices
an increase of +1.0 cm$^{-1}$ in $\omega_e$ and a decrease of 0.00006 \AA\ in
$r_e$. We also performed some calculations with a doubly augmented
cc-pV5Z basis set (i.e. d-AV5Z), and found the results to be essentially
indistinguishable from those with the singly augmented basis set.
Residual imperfections in the electron correlation method can be gauged
from the CAS(8/7)-ACPF $-$ FCI difference with our smallest basis set,
and appear to consist principally of a contraction of $r_e$ by 0.00004 \AA\ and
a decrease in $\omega_e$ by 1.5 cm$^{-1}$. Adding the two sets of differences
to obtain a `best nonrelativistic' set of spectroscopic constants, we obtain
$\nu$=3558.6 cm$^{-1}$, still 3 cm$^{-1}$ above experiment. In both cases,
changes in the anharmonicity constants from the best directly computed 
results are essentially nil.

Scalar relativistic corrections were computed at the CAS(8/7)-ACPF level
with and without the $(1s)$-like electrons correlated, and with a variety
of basis sets. All results are fairly consistent with those obtained at
the highest level considered, CAS(8/7)-ACPF(all)/ACVQZ, namely an
expansion of $r_e$ by about 0.0001 \AA\ and --- most importantly for 
our purposes --- a decrease of $\omega_e$ by about 3 cm$^{-1}$. Effects
on the anharmonicity constants are essentially nonexistent. 

Upon adding these corrections to our best nonrelativistic spectroscopic
constants, we obtain our final best estimates. These lead to $\nu$=3555.44
cm$^{-1}$ for $^{16}$OH$^-$, in excellent agreement with the experimental
result\cite{Ros86} 3555.6057(22) cm$^{-1}$. The discrepancy between computed
(3544.30 cm$^{-1}$)
and observed\cite{Ros86} (3544.4551(28) cm$^{-1}$) values for $^{18}$OH$^-$
is quite similar.  For $^{16}$OD$^-$, we obtain
$\nu$=2625.31 cm$^{-1}$, which agrees to better than 0.1 cm$^{-1}$ with
the experimental value\cite{Reh86} 2625.332(3) cm$^{-1}$. Our computed
bond length is slightly shorter than the observed one\cite{Ros86} for 
OH$^-$, but within the error bar of that for OD$^-$\cite{Reh86}. If we
assume an inverse mass dependence for the experimental
diabatic bond distance and 
extrapolate to infinite mass, we obtain an experimentally derived
Born-Oppenheimer bond distance of 0.96416(16) cm$^{-1}$, in perfect 
agreement with our calculations.

While until recently it was generally assumed that scalar relativistic
corrections are not important for first-and second-row systems, 
it has now been shown repeatedly (e.g.\cite{Bau98,W1,bf3cwb}) that for
kJ/mol accuracy on computed bonding energies, scalar relativistic
corrections are indispensable. Very recently, Csaszar et al.\cite{h2o-rel}
considered the effect of scalar relativistic corrections on the
ab initio water surface, and found corrections on the same
order of magnitude as seen for the hydroxyl anion here. Finally,
Bauschlicher\cite{BauCF4} compared first-order Darwin and mass-velocity
corrections to energetics
(for single-reference ACPF wave functions) with more
rigorous relativistic methods (specifically, Douglas-Kroll\cite{dk}), 
and found that for first-and second-row
systems, the two approaches yield essentially identical results,
lending additional credence to the results of both Csaszar et al.
and from the present work.
(The same author found\cite{Bau99} more significant deviations for third-row
main group systems.) 

Is the relativistic effect seen here in OH$^{-}$ unique to it, or 
does it occur in the neutral first-row diatomic hydrides as well?
Some results obtained for BH, CH, NH, OH, and HF in their respective 
ground states, and using the same method as for OH$^{-}$, are 
collected in Table \ref{rel}. In general, $\omega_{e}$ is slightly
lowered, and $r_{e}$ very slightly stretched --- these tendencies 
becoming more pronounced as one moves from left to right in the 
Periodic Table. The effect for OH$^{-}$ appears to be stronger than
for the isoelectronic neutral hydride HF, and definitely compared
to neutral OH. The excellent agreement ($\pm1$ cm$^{-1}$ on 
vibrational quanta) previously seen\cite{ch} for the first-row 
diatomic hydrides between 
experiment and CCSD(T)/ACV5Z potential curves with an FCI correction 
is at least in part due to a cancellation between the effects of 
further basis set extension on the one hand, and scalar relativistic
effects (neglected in Ref.\cite{ch}) on the other hand. The shape of 
the relativistic contribution to the potential curve is easily understood 
qualitatively:
on average, electrons are somewhat further away from the nucleus in a 
molecule than in the separated atoms (hence the scalar relativistic
contribution to the total energy will be slightly smaller in absolute
value at $r_{e}$ than in the dissociation limit): as one approaches 
the united atom limit, however, the contribution will obviously
increase again. The final result is a slight reduction in both the 
dissociation energy and on $\omega_{e}$.

In order to assist future experimental studies on OH$^-$ and its
isomers, predicted vibrational quanta $G(n)-G(n-1)$ 
are given in Table \ref{Gn} for various isotopic
species, together with some key spectroscopic constants. 
The VBM parameters of the potential are given in Table \ref{vbm}.
The VBM expansion generally
converges quite rapidly\cite{Cox92} and, as found previously for OH, 
parameters $b_5$ and $b_6$ are found to be statistically not significant
and were omitted. 

The VBM expansion requires the insertion of a dissociation energy: 
we have opted, rather than an experimental value, to use our best calculated value
(see next paragraph).

Agreement between computed and observed fundamental frequencies speaks
for itself, as does that between computed and observed rotational constants.
At first sight agreement for the rotation-vibration coupling constants
$\alpha_e$ is somewhat disappointing. However, for $^{16}$OH$^-$ and
$^{18}$OH$^-$, the experimentally derived `$\alpha_e$' actually 
corresponds to $B_1-B_0$, i.e. to $\alpha_e-2\gamma_e+\ldots$. If
we compare the observed $B_1-B_0$ with the computed $\alpha_e-2\gamma_e$
instead, excellent agreement is found. In the case of $^{16}$OD$^-$,
the experimentally derived $\alpha_e$ given is actually extrapolated
from neutral $^{16}$OD: again, agreement between computed and 
observed $B_1-B_0$ is rather more satisfying.

We also note that our calculations validate the conclusion by Lee and Dateo
that the experimentally derived $\omega_e$ and $\omega_ex_e$ for $^{16}$OH
should be revised upward.

\subsection{Dissociation energies of OH and OH$^-$; electron affinity of OH}

This was obtained in the following manner, which is a variant on W2 theory\cite{W1}:
(a) the CASSCF(8/7) dissociation energy using ACVTZ, ACVQZ, and ACV5Z basis sets
was extrapolated geometrically using the geometric formula $A+B/C^n$ first
proposed by Feller\cite{Fel92}; (b) the dynamical correlation component 
(defined at CAS(8/7)-ACPF(all) $-$ CASSCF(8/7)) of the
dissociation energy was extrapolated to infinite maximum angular momentum in the
basis set, $l\rightarrow\infty$ from the ACVQZ ($l$=4) and ACV5Z ($l$=5)
results using the formula\cite{Hal98} $A+B/l^3$; (c) the scalar relativistic 
contribution obtained at the CAS(8/7)-ACPF level was added to the total, as
was the spin-orbit splitting\cite{Moo63} for O$^-$($^2P$). Our final result,
$D_0$=4.7796 eV, is about 0.02 eV higher than the experimental one\cite{Hub79};
interestingly enough, the same is true for the OH radical (computed 
$D_0$=4.4124 eV, observed 4.392 eV). 
In combination with either the experimental electron affinity of oxygen 
atom, EA(O)=1.461122(3) eV\cite{eaO}
or the best computed EA(O)=1.46075 eV\cite{ea}, this leads to 
electron affinities of OH, EA(OH)=1.8283 eV and 1.8280 eV, respectively,
which agree to three decimal places
with the experimental value\cite{Smi97} 1.827611(4) eV. We note that the
experimental $D_e$(OH$^-$) is derived from $D_e$(OH)$+$EA(OH)$-$EA(O),
and that a previous calibration study on the atomization energies of the
first-row hydrides\cite{hydrides} suggested that the experimental $D_e$(OH)
may be too low. While a systematic error in the electronic structure treatment
that cancels almost exactly between OH and OH$^-$ cannot entirely be ruled
out, the excellent agreement obtained for the electron affinity does lend
support to the computed $D_e$ values.

\section{Conclusions}

We have been able to obtain a fully ab initio radial function of spectroscopic
quality for the hydroxyl anion. In order to obtain accurate results for
this system, inclusion of connected quadruple excitations (in a coupled
cluster expansion) is imperative, as is an account for scalar relativistic
effects. Basis set expansion effects beyond $spdfgh$ take a distant third
place in importance. While consideration of connected quadruple excitation
effects and of basis set expansion effects beyond $spdfgh$ would at present be
prohibitively expensive for studies of larger anions, no such impediment would 
appear to exist for inclusion of the scalar relativistic effects (at 
least for one-electron Darwin and mass-velocity terms).

Our best computed EA(OH), 1.828 eV, agrees to three decimal
places with the best available experimental value. Our best computed
dissociation energies, $D_0$(OH$^-$)=4.7796 eV and $D_0$(OH)=4.4124 eV,
suggest that the experimental $D_0$(OH)=4.392 eV (from which the experimental
$D_0$(OH$^-$) was derived by a thermodynamic cycle) may possibly be about 
0.02 eV too low.

One of the purposes of the paper by Lee and Dateo\cite{Lee97} was
to point out to the scientific community, and in particular the 
experimental community, that state-of-the art ab initio methods now have
the capability to predict the spectroscopic constants of molecular anions
with sufficient reliability to permit assignment of a congested spectrum 
from an uncontrolled
environment --- such as an astronomical observation --- on the basis
of the theoretical calculations alone. The present work would appear
to support this assertion beyond any doubt.

\acknowledgments

JM is the incumbent of the Helen and Milton A. Kimmelman Career 
Development Chair.
Research at the Weizmann Institute 
was supported by the Minerva Foundation, Munich, Germany, and
by the {\em Tashtiyot} program of the Ministry of Science (Israel).

\begin{table}
\caption{\label{npart}Computed total energy (hartree), bond distance (\AA), harmonic frequency (cm$^{-1}$)
and anharmonicity constants (cm$^{-1}$) of $^{16}$OH$^-$ using the cc-pVDZ+sp(O) basis set
as a function of the electron correlation method}
\begin{tabular}{ldddddd}
 &  $E_e$ & $r_e$ & $\omega_e$ & $\omega_ex_e$ & $\omega_ey_e$ & $\omega_ez_e$\\
\hline
FCI              & -75.623457 & 0.97503 & 3701.7 & 96.65 & 0.454 & -0.024\\
CCSD             & -75.616478 & 0.97209 & 3747.1 & 95.28 & 0.537 & -0.010\\
CCSD(T)          & -75.622380 & 0.97442 & 3711.6 & 96.45 & 0.401 & -0.031\\
CC5SD(T)         & -75.621379 & 0.97428 & 3709.5 & 97.74 & 0.367 & -0.025\\
CCSDT            & -75.622656 & 0.97449 & 3709.9 & 96.37 & 0.465 & -0.023\\
CCSD(TQ)         & -75.621660 & 0.97467 & 3703.1 & 98.17 & 0.352 & -0.024\\
CCSD+TQ*         & -75.621473 & 0.97463 & 3702.8 & 98.48 & 0.337 & -0.023\\
approx. CCSDT(Q) & -75.622937 & 0.97488 & 3703.5 & 96.78 & 0.452 & -0.022\\
approx. CCSDT+Q* & -75.622750 & 0.97484 & 3703.2 & 97.10 & 0.438 & -0.020\\
CAS(8/7)-ACPF    & -75.623089 & 0.97499 & 3703.2 & 96.60 & 0.455 & -0.023\\
CAS(8/7)-AQCC    & -75.622147 & 0.97500 & 3702.9 & 96.54 & 0.456 & -0.029\\
CAS(8/8)-ACPF    & -75.623084 & 0.97501 & 3703.0 & 96.66 & 0.444 & -0.024\\
CAS(8/8)-AQCC    & -75.622669 & 0.97493 & 3704.2 & 96.59 & 0.443 & -0.024\\
\end{tabular}

\end{table}

\begin{table}
\caption{\label{1part}Computed bond distance, harmonic frequency, anharmonicity constants, and Dunham correction to harmonic frequency
for $^{16}$OH$^-$ as a function of basis set and electron correlation method. All data in cm$^{-1}$ except $r_e$ (\AA)}
\squeezetable
\begin{tabular}{lllddddddd}
Corr. method & basis set & 1s corr? & $r_e$ & $\omega_e$ & $\omega_ex_e$ & $\omega_ey_e$ & $\omega_ez_e$ & $Y_{10}-\omega_e$ & $\nu$\\
\hline
CAS(8/7)-ACPF & aug'-cc-pVTZ &no & 0.96776 & 3725.01 & 92.738 & 0.3623 & -0.0566 & -0.37 & 3540.07\\
CAS(8/7)-ACPF & aug'-cc-pVQZ &no & 0.96517 & 3742.24 & 93.610 & 0.3855 & -0.0068 & -0.24 & 3556.00\\
CAS(8/7)-ACPF & aug'-cc-pVQZ & no+REL & 0.96528 & 3739.00 & 93.564 & 0.3881 & -0.0066 & -0.24 & 3552.86\\
CAS(8/7)-ACPF & aug'-cc-pV5Z &no & 0.96476 & 3745.58 & 93.856 & 0.4968 & -0.0192 & -0.14 & 3559.24\\
CCSD(T) & aug'-cc-pVTZ &no & 0.96741 & 3733.55 & 91.987 & 0.3284 & -0.0524 & -0.40 & 3549.99\\
CCSD(T) & aug'-cc-pVQZ &no & 0.96486 & 3750.37 & 92.948 & 0.3474 & -0.0121 & -0.27 & 3565.28\\
CCSD(T) & aug'-cc-pV5Z &no & 0.96456 & 3751.56 & 93.183 & 0.4643 & -0.0227 & -0.17 & 3566.42\\
\hline
CAS(8/7)-ACPF & AVTZ &no & 0.96809 & 3716.44 & 92.083 & 0.2144 & -0.0133 & -0.42 & 3532.49\\
CAS(8/7)-ACPF & AVQZ &no & 0.96551 & 3737.30 & 93.868 & 0.4277 & -0.0034 & -0.19 & 3550.75\\
CAS(8/7)-ACPF & AV5Z &no & 0.96488 & 3744.47 & 93.816 & 0.5236 & -0.0157 & -0.13 & 3558.33\\
CCSD(T) & AVTZ &no & 0.96781 & 3723.56 & 91.345 & 0.1745 & -0.0188 & -0.46 & 3540.88\\
CCSD(T) & AVQZ &no & 0.96520 & 3745.61 & 93.159 & 0.3900 & -0.0107 & -0.22 & 3560.29\\
CCSD(T) & AV5Z &no & 0.96472 & 3749.39 & 93.193 & 0.4966 & -0.0291 & -0.15 & 3564.32\\
CCSD(T) & d-AV5Z &no & 0.96476 & 3749.31 & 93.079 & 0.4900 & -0.0283 & -0.16 & 3564.45\\
CCSD(T) & AV6Z &no & 0.96466 & 3750.41 & 93.237 & 0.4839 & -0.0214 & -0.14 & 3565.26\\
CCSD(T) & AVTZuc &no & 0.96734 & 3724.84 & 92.600 & 0.4875 & -0.0734 & -0.39 & 3540.46\\
CCSD(T) & AVQZuc &no & 0.96522 & 3744.72 & 93.044 & 0.4081 & -0.0219 & -0.27 & 3559.58\\
CCSD(T) & AV5Zuc &no & 0.96473 & 3749.21 & 93.243 & 0.4435 & -0.0103 & -0.16 & 3563.95\\
\hline
CAS(8/7)-ACPF & ACVTZ &no & 0.96789 & 3713.45 & 91.642 & 0.2137 & 0.0000 & -0.41 & 3530.45\\
CAS(8/7)-ACPF & ACVQZ &no & 0.96558 & 3735.72 & 93.894 & 0.4219 & -0.0130 & -0.23 & 3549.01\\
CAS(8/7)-ACPF & ACV5Z &no & 0.96501 & 3740.66 & 94.081 & 0.4691 & -0.0005 & -0.14 & 3553.87\\
CCSD(T) & ACVTZ &no & 0.96768 & 3718.89 & 91.145 & 0.1639 & -0.0044 & -0.45 & 3536.66\\
CCSD(T) & ACVQZ &no & 0.96525 & 3744.90 & 93.038 & 0.3867 & -0.0191 & -0.26 & 3559.73\\
CCSD(T) & ACV5Z &no & 0.96472 & 3749.22 & 93.225 & 0.4361 & -0.0101 & -0.17 & 3563.96\\
CAS(8/7)-ACPF & ACVTZ &yes& 0.96725 & 3714.74 & 92.017 & 0.1855 & -0.0035 & -0.43 & 3530.86\\
CAS(8/7)-ACPF & ACVQZ &yes& 0.96468 & 3741.86 & 94.110 & 0.4205 & -0.0129 & -0.23 & 3554.71\\
CAS(8/7)-ACPF & ACV5Z &yes& 0.96410 & 3746.51 & 94.317 & 0.4682 & 0.0009 & -0.14 & 3559.26\\
CCSD(T) & ACVTZ &yes& 0.96688 & 3725.04 & 91.122 & 0.1509 & -0.0022 & -0.46 & 3542.81\\
CCSD(T) & ACVQZ &yes& 0.96435 & 3751.76 & 93.151 & 0.3929 & -0.0202 & -0.26 & 3566.37\\
CCSD(T) & ACV5Z &yes& 0.96378 & 3756.27 & 93.347 & 0.4427 & -0.0088 & -0.17 & 3570.80\\
CAS(8/7)-ACPF & ACVQZ all & yes+REL & 0.96478 & 3738.69 & 94.098 & 0.4193 & -0.0102 & -0.24 & 3551.57\\
$\Delta$REL &  &  & 0.00010 & -3.17 & -0.012 & -0.0012 & 0.0027 & -0.01 & -3.14\\
best calc. &  &  & 0.96417 & 3742.87 & 94.404 & 0.4527 & 0.0100 & -0.14 & 3555.44\\
\end{tabular}

The suffix ``+REL'' indicates inclusion of scalar relativistic (Darwin and mass-velocity)
effects obtained as expectation values for the wave function indicated.
\end{table}

\begin{table}
    \caption{\label{rel}Effect of scalar relativistic contributions 
    on the bond lengths (\AA) and harmonic frequencies (cm$^{-1}$) 
    of the AH (A=B--F) diatomics. All calculations were carried out 
    at the CAS($2\sigma3\sigma4\sigma1\pi$)-ACPF/ACVQZ level with all
    electrons correlated}
\begin{minipage}{4in}
    \begin{tabular}{ldd}
       & $\Delta r_e$ & $\Delta\omega_e$\\
\hline
BH     &  -0.00001  &  -0.57 \\
CH     &  +0.00001  &  -1.08 \\
NH     &  +0.00003  &  -1.77 \\
OH     &  +0.00004  &  -2.35 \\
HF     &  +0.00005  &  -2.80 \\
OH$^-$ &  +0.00010  &  -3.14 \\
\end{tabular}
\end{minipage}

Effects on the anharmonicity constants are negligible.

\end{table}

\begin{table}
\caption{\label{vbm}Parameters for the VBM representation, eq. (1), 
obtained from our best potential. $D_e$, $r_e$ are in cm$^{-1}$ and \AA, respectively;
the remaining parameters are dimensionless}
\begin{minipage}{3in}
\begin{tabular}{ld}
$D_e$  & 40398.7079\\
$r_e$  &   0.964172\\
$\beta$&   2.128977\\
$b_1$  & -0.047181\\
$b_2$  &  0.022371\\
$b_3$  & -0.0070906\\
$b_4$  &  0.0018429\\
\end{tabular}
\end{minipage}
\end{table}

\begin{table}
\caption{\label{Gn}Spectroscopic constants and band origins (in cm$^{-1}$) of different isotopomers
of the hydroxyl anion obtained from our best potential}
\squeezetable
\begin{tabular}{lccccccc}
                            & \multicolumn{2}{c}{$^{16}$OH$^-$} & \multicolumn{2}{c}{$^{16}$OD$^-$} & \multicolumn{2}{c}{$^{18}$OH$^-$} & $^{18}$OD$^-$\\
                            & calc & obsd$^a$ & calc & obsd$^b$& calc & obsd$^a$& calc \\
\hline
$Y_{00}$                    &    2.38     &  &    1.26     &  &    2.36     &  &    1.25    \\   
$Y_{10}\approx\omega_e$     & 3742.72     & 3738.44(99)$^c$ & 2724.79     & 2723.5(10) & 3730.35     &  & 2707.77    \\
$-Y_{20}\approx\omega_exe$  &   94.298    & 91.42(49)$^c$ &   49.979    & 49.72(50) &   93.676    &  &   49.357   \\
$Y_{30}\approx\omega_eye$   &    0.4686   &  &    0.1808   & 0.38(15) &    0.4639   &  &    0.1774  \\
$Y_{01}\approx B_e$         &   19.126021 & 19.12087(37) &   10.136936  & 10.13599(30) &18.999788 & 18.99518(49) &   10.010698\\
$-Y_{11}\approx\alpha_e$    &    0.779874 & 0.77167(13) &    0.300914 & 0.3043(5) &    0.772165 &0.76409(16)  &    0.295310\\
$Y_{21}\approx\gamma_e$     &    0.003913 &  &    0.001099 &  &    0.003861 &  &    0.001072\\
$\alpha_e$-2$\gamma_e$      &    0.772048 & 0.77167(13)&0.298716&0.2984(3)&0.764443&0.76409(16)&0.293166\\
$-Y_{02}\approx D_e$        &    0.001998 & 0.001995(6) &    0.000561 & 0.000559(2)$^d$ &    0.001972 & 0.000031(2) &    0.000547\\
$Y_{12}\approx\beta_e$      &    0.000032 & 0.000032(2) &    0.000006 & 0.000008(2) &    0.000031 & 0.000031(2)  &    0.000006\\
 ZPVE                       & 1850.23 && 1351.19 && 1844.18 && 1342.81\\ 
G(1)-G(0)                   & 3555.63 & 3555.6057(22) & 2625.42 & 2625.332(3) &3544.49 & 3444.4551(28)& 2609.63\\
G(2)-G(1)                   & 3371.17 && 2527.06 && 3361.24 && 2512.49\\
G(3)-G(2)                   & 3189.42 && 2429.75 && 3180.66 && 2416.38\\
G(4)-G(3)                   & 3010.39 && 2333.49 && 3002.78 && 2321.29\\
G(5)-G(4)                   & 2834.11 && 2238.28 && 2827.63 && 2227.23\\
G(6)-G(5)                   & 2660.70 && 2144.12 && 2655.31 && 2134.21\\
G(7)-G(6)                   & 2490.31 && 2051.03 && 2485.97 && 2042.24\\
\end{tabular}

The Dunham constants $Y_{mn}$ include higher-order corrections to the mechanical
spectroscopic constants (like $\omega_e, \omega_ex_e$) as obtained from the 
potential function.

(a) Ref.\cite{Ros86}. Uncertainties in parentheses correspond to two standard deviations.

(b) Ref.\cite{Reh86}. Uncertainties in parentheses correspond to three standard deviations.

(c) LD proposed $\omega_e$=3741.0(14) and $\omega_ex_e$=93.81(93) cm$^{-1}$, obtained
by mass scaling of the $^{16}$OD$^-$ results, as more reliable.

(d) From observed $D_0$ and $D_1$ in Ref.\cite{Reh86}.

\end{table}

\begin{figure}
    \caption{Deviation from the FCI potential curve of OH$^-$ for 
    different electron correlation methods}
    \vspace*{0.5in}
    \includegraphics[scale=0.90]{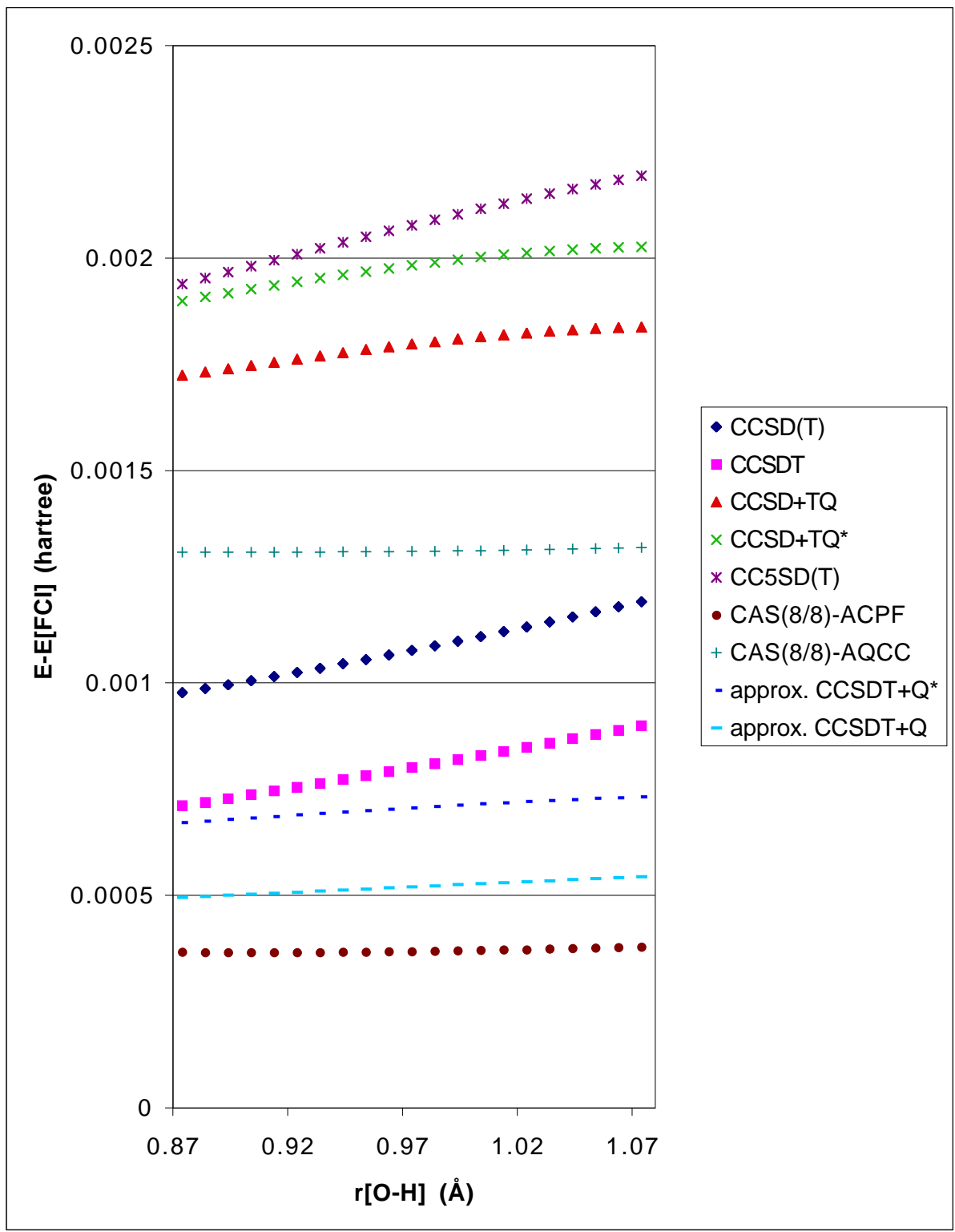}
    \end{figure}

\end{document}